# Unexpected colloid-like supernatant from s liquid-phase ball-milling graphite using miscible solutions as solvents: a failure analysis

Ling Sun
Hokkaido University, Sapporo 060-0810, Japan

## Abstract

Ball-milling graphite was conducted in miscible solutions with the purpose to exfoliate graphene. Colloid-like stable supernatants were unexpectedly obtained. Followed were the characterizations with Scanning electronic microscopy, Thermogravimetric analysis, X-ray diffraction and Elemental dispersive X-ray mapping to recognize them. As a result, strongly against the initial judgments, the components of colloid were mainly impurities of mixture of zirconia, silicate and yttrium oxide, other than nano-graphite.

## Introduction

We attempted to isolate graphene sheets from graphite by using various methods. On one hand, Wet oxidation-reduction method has been ever tried. As a result, we developed one modified Hummers method for mass production of graphene oxide with improved productivity and safety [1]. On the other, we also desire of highly effective way to obtain graphene directly from physicochemical exfoliation. Such methods need to overcome the interlayer van der Waals forces by virtue of solvent/molecular -graphite interactions that are equivalent to that surface energy [2, 3]. Planetary milling, wherever in solid or liquid phases, that executes mainly shear forces, is recently frequently reported to prepare graphene sheets with kinds of intercalating agents [3, 4]. Most of them are to focus on improvement of exfoliation efficiency. Herein we tried kinds of mixed solution to mill graphite for a certain time to study the possibility of graphene preparation. We found in our ball milling system the graphite itself reached smaller sizes, whilst there produced white colloid-like supernatants, which had nice solubility and stability. We initially doubted they might be graphite nanoparticles, but as experiments proved, most of them were impurities.

## Materials and methods

**Colloid-like Supernatant from Ball milling of graphite at different media**

Graphite was subjected to ball milling using a planetary milling setup (SEIWA GIKEN Co. Ltd., RM-10). Typically, 2 g graphite (SP-1, Bay Carbon) powder and beads ($Y_2O_3$-$ZrO_2$, 6 g/cm$^3$, 2.3 Kg, 5 mm in diameter) were put into a ceramic vessel with an inner volume of ~500 ml. Afterwards, 300-ml mixed solution was added. The solution with a fixed total volume (300ml) was $NH_3$ (25wt%, 30ml) + Water, or Acetone (50ml) + Water, or GO (0.69wt%, ~10 ml) + Water. The milling worked at 220 rpm to ensure the shear stress and was lasted for 24 h at room temperature (~398K). After that, as-obtained mixture was placed quietly to let large graphite flakes subside and then suffered a vacuum filtration to purify the supernatant (filter paper: Comfort service style 200, Nippon paper group).

**Syntheses of GO**

GO was also obtained from natural graphite (SP-1, Bay Carbon) using a modified Hummers and Offeman method [1]. In a typical treatment, potassium permanganate (15 g) and expanded graphite (5 g) were stirred to be homogeneous. Next, in a bottom-round flask (500 ml) with ice-water bath, concentrated sulfuric acid (98 %, 100 ml) was added with continuously stirring until a uniform liquid paste formed. Then removed the water bath and the stirring was continued and stopped until a foam-like intermediate spontaneously formed (around 30 min) with a large volumetric expansion. Afterwards, deionized water (400 ml) was added. Next, in a 90 ℃ water bath for one hour a homogeneous suspension was obtained dark yellow in color. It was then filtered and was subjected with repeated washing and centrifugation (10000 rpm, 2 h per cycle) to remove impurities.

**Characterizations**

The characterization involved Scanning electro-microscopy (SEM, JSM-6300, JOEL), Thermogravimetric analysis (TGA, TG/DTA 6200, SII Exstar6000, heating rate of 5 K per minute in air) and X-ray diffraction (XRD, Rigaku Denki RINT 2000, X-ray $\lambda_{Cu\,k\alpha}$= 0.154 nm), STEM (Hitachi HD-2000 Scanning Transmission Electron Microscope, accelerating voltage 200 KV equipped with an Energy Dispersive X-ray (EDX) detector).

# Results and discussion

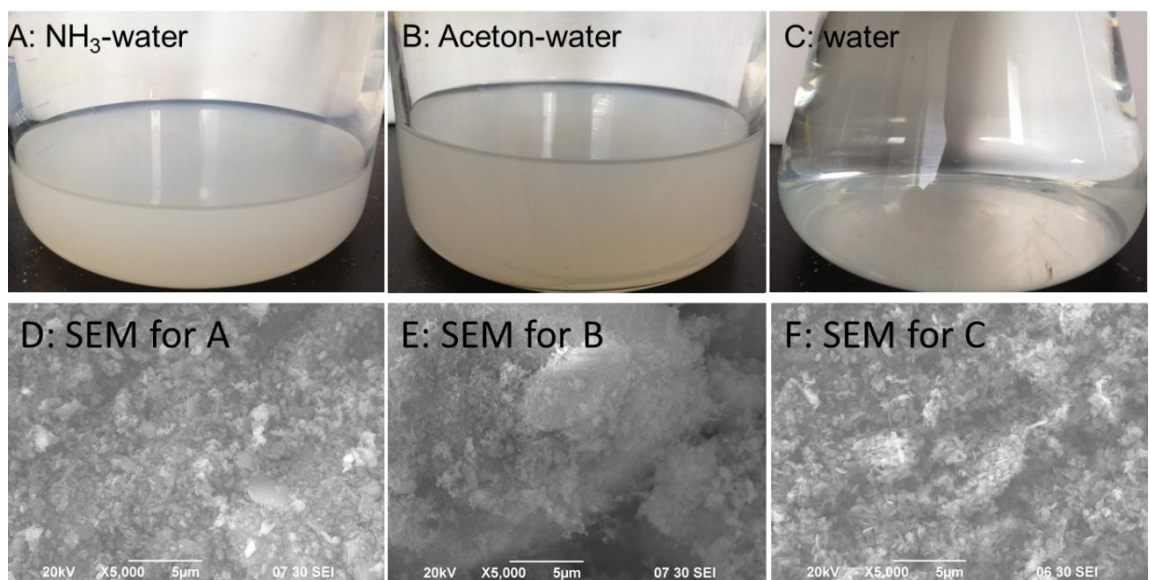

Fig.1. Supernatant photos of ball-milling graphite at 220rpm for 24 h and their respective SEM morphology images.

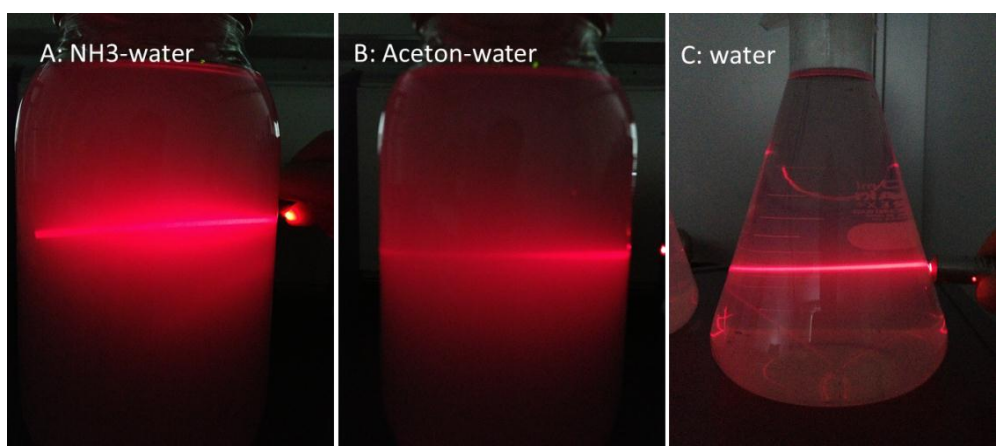

Fig.2. Photos of Tyndal phenomena of supernatants of ball-milling graphite (laser wavelength 650nm).

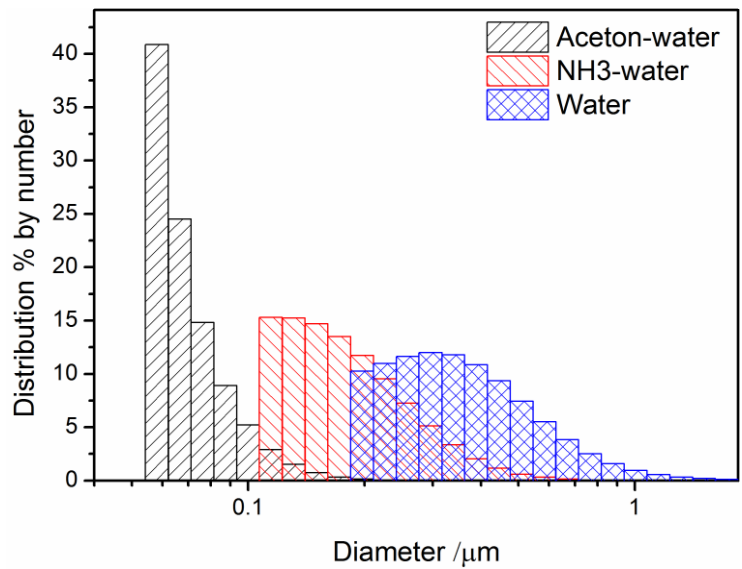

Fig. 3. DLS analysis on size distribution of the colloidal particle.

Ball milling of graphite at different media was conducted. As shown in Fig. 1, after filtration, we obtained white colloid-like solutions. Even in the case water being the only solvent involved, we still recognized a slightly cloudy solution. With several-day standing, a little white precipitate formed and deposited at the bottom, indicative of the colloid particles being somewhat not homogeneous in the supernatant. Even though, the supernatants were able to be colloidal-like stable over several months (Fig. 2). With drying them at a 313 K oven, the morphologies were then observed under SEM (Fig.1 d, e, f). Particles at distinctly different sizes were seen. Amongst them, we did be aware that the particles in the case of acetone were the smallest in average while the sizes in the case of water were the largest (not shown). And this was consistent to the result from DLS analysis (the median diameter order: 61.1 nm (acetone) < 157.6 nm ($NH_3$) < 314.1 nm (water), see Fig. 3).

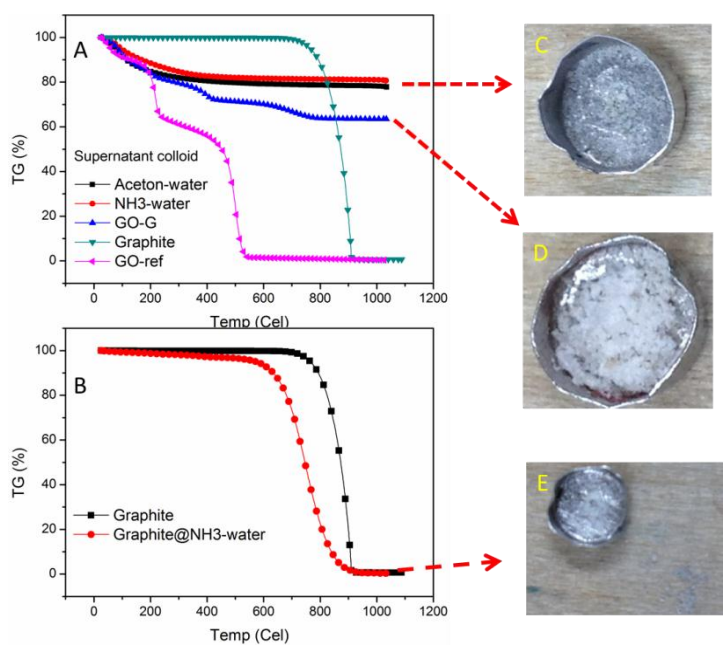

Fig.4. TGA analysis of dry supernatant matter at different media and their final products.

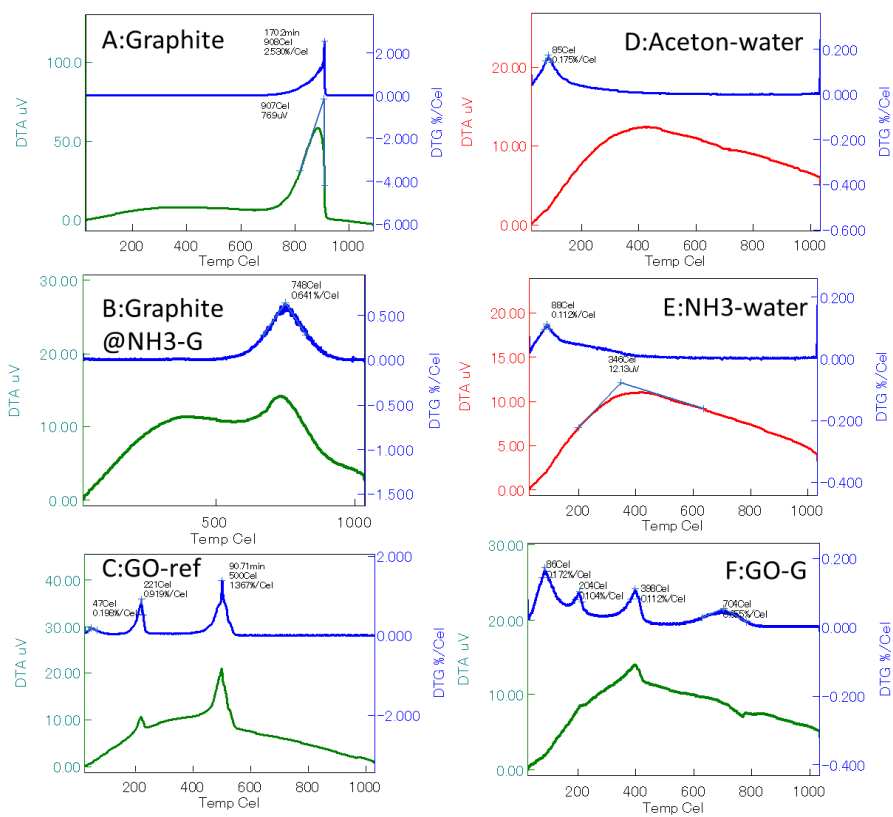

Fig.5. DTG/DTA curves.

| Samples | DTG peak (Position /K, Value /%/K) |
|---|---|
| Graphite | [a] 1181, 2.53 |
| Graphite@NH$_3$-water | [a] 1020, 0.641 |
| Aceton-water | 358, 0.175 |
| NH$_3$-water | 361, 0.112 |
| GO-Ref | 320, 0.198; [b] 494, 0.919; [b] 773, 1.367 |
| GO-G | 359, 0.172; [b] 477, 0.104; [b] 671, 0.112; [a] 977, 0.055 |

Tab. 1. DTG information. Note: [a] represents the peaks of graphite combustion indicative of the maximum decomposition rates; [b] represents that characteristic peaks for GO.

We further used TG to analyze the samples annealed in air. Of note herein, we also tried milling the graphite in GO solution. In fact, similar to the foregoing trials, graphite appeared to have the same trend, that is, sizes decreased and meanwhile, smaller graphite flakes turned out to be better dispersed as compared to the raw graphite with relatively larger sizes. This is easily understandable. GO has in nature amphiphilic property. It has been demonstrated capable of dispersing a variety of π-conjugated carbon materials already [5]. Through such wet route, such all-carbon composites became much promising at mass scale. As-obtained graphite dispersion after removing the precipitated larger flakes was then dried and also annealed in air. Graphite and GO were at the same used as references. As shown in Fig. 4a, reference graphite demonstrated well thermal stability at less than 1173K, followed with such full burning as to that little ash remained. Reference GO was much alike to what we ever annealed [6]. Typically it is composed of the adsorbed/crystallized water release, instable functional groups removal, and the carbon network decomposition [6]. As compared and unexpectedly, for the tested samples, there remained large proportions (at least over 60%) as impurities, which was called as white residues. We initially thought they might be carbonate salt, and then put them in 1wt% $H_2SO_4$. However they kept incompletely dissolved, which proved that part of them was surely the carbonate salt. As to the remaining, we assigned them as incombustible materials, like metal oxides. For the graphite in the case of GO (Fig. 4b), after annealing, a little "white residue" was also formed, similar to its supernatant result, but much less on amount.

Fig. 5 and Tab. 1 show further differentiated TG (DTG) results of TGA and peak information respectively. These would help better understand the thermal features of all samples in air. By comparing Figs. 5a, d, and e, graphite features a typical peak at 1181K, while after milling, sizes being smaller, the characteristic peak values deceased to 1020 K (graphite@$NH_3$-water), and 977 K (GO-G). Such peak, however almost lost in the cases of acetone and $NH_3$, which proved (1) much less portion of graphite than that of impurities, (2) so smaller graphite as to hardly produce their signal (the lack of weight loss at over 1073 K as shown in Fig. 4).

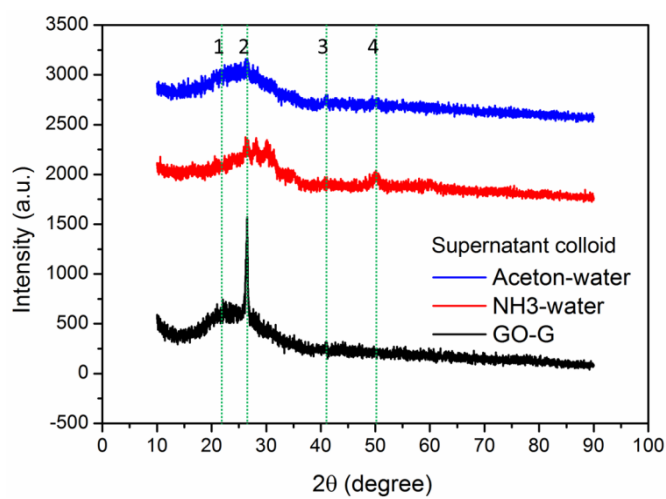

Fig.6. XRD patterns of dry supernatant matter at different media.

And the above conclusion could be also verified by the XRD regarded as a lossless characterization (Fig. 6). We simply compared the three samples' XRD patterns without involving graphite and GO as references. As a matter of fact, it is clearly shown that the typical graphite peak (line 2) at 26° (001, interlayer spacing ~0.34 nm) in all patterns. Differently, intensities for the samples in the cases of acetone and $NH_3$ were obviously much lower than that in the GO-G case. Line 1 illustrates a typical broadened shifted peak of GO at ~22.6° (001, interlayer spacing ~0.4 nm), while it did not show in the other groups. Such two typical peaks got partially overlapped in the GO-G pattern, while for others, only broadened graphite peaks existed. There were actually more peaks in the case of acetone and $NH_3$. With excluding the possibility of being graphite, the impurities should be responsible.

It actually was interesting our ball milling system had introduced a large amount of impurities into supernatant. To date, to our knowledge, no one has ever reported such negative phenomenon. In most cases, they preferred using sole-component liquor/solid solvent, like acetone, ethanol, formamide [3], ammonia borane [4]. Miscible solutions used as milling solvent probably was the first reported. Similarly to otherwise reports, graphite flaked were milled to become smaller, and even thinner, approaching tens of nano-meters with the 24-h milling. Such results were actually not what this paper concentrated at and therefore omitted. As above mentioned, at the experimental beginning, we expected much of nano-graphite particles that constituted the main component. But the truth was what we had were in fact dominantly impurities. Checking our system, we temporarily reasoned our failure being the result of using $Y_2O_3$-$ZrO_2$ beads in a ceramic vessel.

To clarify the truth, EDX was used to obtain the mapping images of element C, Si, O, Zr and Y. As shown in Fig. 7, amazingly but highly conforming to the above discussion, impurity elements Si, O, and Zr had extremely higher distribution over that of C and Y. Si was introduced from the ceramic vessel, while O and Zr to high extent were from the milling beads and air atmosphere inside, as well as the Y. As-obtained supernatant was actually of the collections of these impurities of oxidized matter, and some carbonaceous particles.

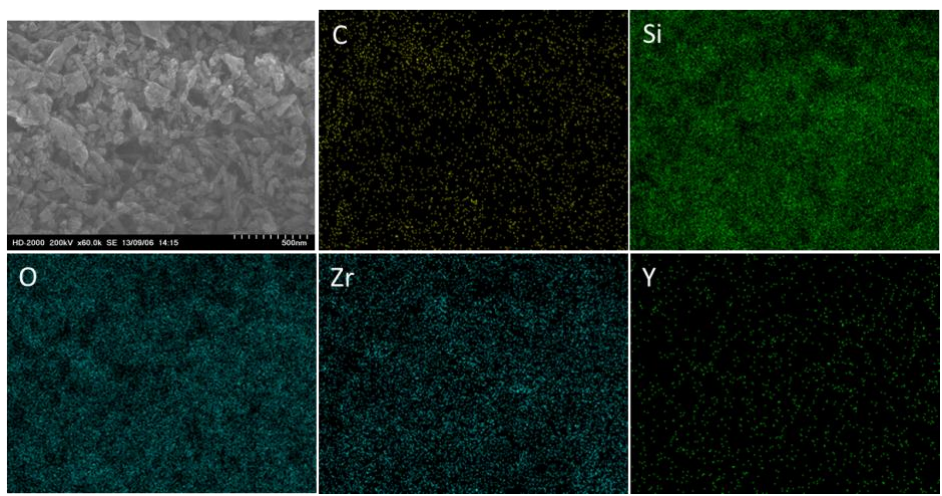

Fig.7. SEM image and corresponding EDX elemental mapping images of dry supernatant matter.

## Conclusion

Ball-milling graphite at different media ($NH_3$/water, Acetone/water, GO solution, etc) was conducted with the purpose of exfoliating graphite into graphene sheet. Accidentally we found the supernatant being much cloudy and long-term colloid-like stable. Through characterizations it was found the main component was comprised of impurities, other than graphite. In other words, massive obtaining of graphene from supernatant via proposed wet method under mentioned conditions is much unavailable.

## References

[1]  Sun L, Fugetsu B. Mass production of graphene oxide from expanded graphite. Mater Lett 2013;109(0):207-10.
[2]  Hernandez Y, Nicolosi V, Lotya M, Blighe FM, Sun ZY, De S, et al. High-yield production of graphene by liquid-phase exfoliation of graphite. Nat Nanotechnol 2008;3(9):563-8.
[3]  Zhao W, Wu F, Wu H, Chen G. Preparation of Colloidal Dispersions of Graphene Sheets in Organic Solvents by Using Ball Milling. J Nanomater 2010;2010:1-5.
[4]  Liu L, Xiong Z, Hu D, Wu G, Chen P. Production of high quality single- or few-layered graphene by solid exfoliation of graphite in the presence of ammonia borane. Chem Commun 2013;49(72):7890-2.
[5]  Kim J, Cote LJ, Huang J. Two dimensional soft material: new faces of graphene oxide. Acc Chem Res 2012;45(8):1356-64.
[6]  Sun L, Yu H, Fugetsu B. Graphene oxide adsorption enhanced by in situ reduction with sodium hydrosulfite to remove acridine orange from aqueous solution. J Hazard Mater 2012;203-204(0):101-10.